\documentclass[12pt,epsf,amstex]{article}   
\usepackage[utf8]{inputenc}   
\usepackage[T1]{fontenc}
\usepackage{amsmath}
\usepackage{amssymb}

\usepackage{epsfig}

\addtocounter{secnumdepth}{1}
\begin{document}

\newcommand{\betastar}{\beta^\star}
\newcommand{\tbeta}{\tilde{\beta}}

\newcommand{\calHDG}{{\calH}}
\newcommand{\caltHDG}{{\tilde{\calH}}}
\newcommand{\calH}{{\cal H}}
\newcommand{\calN}{{\cal N}}

\newcommand{\corr}{g}
\newcommand{\corrstar}{g^\star}
\newcommand{\corrstarTh}{g^\star_{\infty}}
\newcommand{\Corr}{G}

\newcommand{\kun}{k_1}
\newcommand{\kdeux}{k_2}

\newcommand{\uc}{u^\textrm{C}}
\newcommand{\Uc}{U^\textrm{C}}
\newcommand{\ustar}{u^\star}
\newcommand{\Ustar}{U^\star}

\newcommand{\vstar}{v^\star}
\newcommand{\hv}{\widehat{v}}
\newcommand{\tv}{\widetilde{v}}
\newcommand{\htv}{\widehat{\tv}}
\newcommand{\tvstar}{\widetilde{\vstar}}

\newcommand{\Vstar}{V^\star}

\newcommand{\vnnb}{v^\textrm{nn}}
\newcommand{\hvnnb}{\widehat{\vnnb}}
\newcommand{\tvnnb}{\widetilde{v^\textrm{nn}}}
\newcommand{\htvnnb}{\widehat{\widetilde{\vnnb}}}
\newcommand{\tvnnbTh}{\widetilde{\vnnb_{\infty}}}

\newcommand{\hh}{\widehat{h}}
\newcommand{\hV}{\widehat{V}}
\newcommand{\hVstar}{\widehat{V^\star}}
\newcommand{\tV}{\widetilde{V}}
\newcommand{\htV}{\widehat{\tV}}
\newcommand{\tw}{\widetilde{w}}
\newcommand{\htw}{\widehat{\tw}}
\newcommand{\tW}{\widetilde{W}}

\newcommand{\dd}{\mbox{d}}
\newcommand{\ee}{\mbox{e}}
\newcommand{\ii}{{\rm i}}
\newcommand{\veck}{\mathbf{k}}
\newcommand{\veci}{\mathbf{i}}
\newcommand{\vecr}{\mathbf{r}}
\newcommand{\thalf}{\tfrac{1}{2}}

\newcommand{\beq}{\begin{equation}}
\newcommand{\eeq}{\end{equation}}
\newcommand{\bea}{\begin{eqnarray}}
\newcommand{\eea}{\end{eqnarray}}
\def\lsim{\:\raisebox{-0.5ex}{$\stackrel{\textstyle<}{\sim}$}\:}
\def\gsim{\:\raisebox{-0.5ex}{$\stackrel{\textstyle>}{\sim}$}\:}

\thispagestyle{empty}
\title{\Large {\bf New duality relation for the Discrete Gaussian 
SOS model on a torus}\\[2mm] 
}

\author{F.~Cornu$^1$, H.J.~Hilhorst$^2$ and M.~Bauer$^3$\\[3mm]
$^{1,2}$ {\small Université Paris-Saclay, CNRS} \\
$^1$ {\small Laboratoire de Physique Théorique et Modèles Statistiques, 91405
  Orsay, France}\\
$^2$ {\small Laboratoire Irène Joliot-Curie, 91405 Orsay, France}\\
$^3$ {\small Université Paris-Saclay, CNRS, CEA} \\ 
{\small  Institut de Physique Théorique, 91191 Gif-sur-Yvette, France}\\ 
{\small and PSL Research University, CNRS}\\
{\small École normale supérieure, Département de mathématiques et applications, 75005 Paris, France}
\\}

\maketitle
\begin{small}
\begin{abstract}

We construct a new duality for 
two-dimensional Discrete Gaussian models. It is based on 
a known one-dimensional duality and on
a mapping, implied by the Chinese remainder theorem,
between the sites of an $N\times M$ torus and those of a ring of $NM$ sites.
The duality holds for an arbitrary translation-invariant 
interaction potential $v(\mathbf{r})$ between the height variables on the torus.
It leads to pairs $(v,\widetilde{v})$ of mutually dual potentials and to a
temperature inversion according to $\widetilde{\beta}=\pi^2/\beta$. When  $v(\mathbf{r})$ is isotropic, duality renders an anisotropic $\widetilde{v}$.
This is the case, in particular, for the  potential  that is dual to
an isotropic nearest-neighbor potential.
In the thermodynamic limit this dual potential  is shown to decay 
with distance according to an inverse square law with a quadrupolar angular dependence.
There is a single 
pair of self-dual potentials $v^\star=\widetilde{v^\star}$.
At the self-dual temperature $\beta^\star=\widetilde{\beta^\star}=\pi$ 
the height-height correlation can be calculated explicitly;
it is anisotropic and diverges logarithmically with distance.

\end{abstract}
\end{small}
\vspace{12mm}


\section{Introduction}
\label{sec:Introduction}

The Discrete Gaussian (DG) model  is  a particular lattice model
belonging to the class of the so-called Solid-on-Solid (SoS) models which
aim to describe the fluctuations of a crystal surface.  
The most usual versions of SoS models are two-dimensional. In such a model 
a surface is described as a collection of integer-valued height variables 
$\{h_{\veci}\}$
associated with the sites $\veci$ of a two-dimensional (2D) lattice.
The  interaction between two height variables  $h_{\veci}$ and $h_{\veci^\prime}$
is some function of their difference $|h_{\veci}-h_{\veci^\prime}|$,
and in the case of the DG model it is a simple quadratic form.
When $v(\vecr)$ represents an isotropic  nearest neighbor coupling, 
the  DG model is dual
to the 
XY model in its Villain version \cite{Villain75}, and therefore it undergoes a phase 
transition in the Kosterliz-Thouless universality class. 
This phase transition has been the main motivation for the interest 
in this short-ranged two-dimensional DG model.

The DG Hamiltonians of interest to us in this work take the form
\beq
 \calH = \frac{1}{2} \sum_{\veci}\sum_{\veci^\prime} 
v(\veci-\veci^\prime) (h_{\veci}-h_{\veci^\prime})^2,
\label{HDGd}
 \eeq 
where the coupling constants $v(\vecr)$  constitute a translation-invariant pair potential.
We may impose without loss of generality the symmetry 
$v(\vecr)=v(-\vecr)$
under parity transformation.
We consider a toroidal
lattice of $N\times M$ sites, which for $M=1$ includes also the 
one-dimensional (1D) case.
The partition function associated with Hamiltonian (\ref{HDGd})  reads
$
\sideset{}{'}{\sum}_{\{h_\veci\}}
\exp[{-\beta{\calHDG}}]
$
in which the $h_\veci$ are summed over all integer values except for the
condition, indicated by the prime on the summation sign, 
that one height, say $h_{\veci_0}$, should be kept fixed, say $h_{\veci_0}=0$. 
This ``{global} gauge'' condition 
eliminates a trivial
infinite factor in the partition function, which is due to $\calH$ being invariant under the global
translation $h_\veci \mapsto h_\veci+h$.
After this trivial factor has been removed, the only further
condition on $v(\vecr)$ is that 
(\ref{HDGd}) define a positive definite quadratic form.

The one-dimensional DG model with arbitrary interaction potential $v(r)$
at inverse temperature $\beta$
was studied by  Kjaer and Hilhorst (KH) \cite{KjaerHilhorst82},
who found that it is dual to another such model but with a dual
potential $\tv$ 
and a dual  inverse  temperature $\tbeta=\pi^2/\beta$. Whereas in general $v
\neq \tv$, there is a unique and explicitly known
self-dual potential
$\vstar$ for which $\vstar = \tvstar$.
In the thermodynamic limit $N\to \infty$ the self-dual potential
$\vstar(r)$ tends toward $1/[\pi (r^2-\tfrac{1}{4} )]$. 
The temperature $\betastar$ such that  $\betastar=\widetilde{\betastar}$ is  a candidate for a
critical temperature of this model.

In this work we combine the one-dimensional
KH results with a mapping
between one- and two-dimensional lattices that occurs in number theory
in the context of the 
Chinese remainder theorem.
This theorem suggests to represent the one-dimensional ring lattice 
geometrically as a helix
wound around the two-dimensional torus in such a way that the helix returns to its origin after having passed through all sites on the torus.
The theorem requires that $N$ and $M$ be coprime, that is, have no common
prime factor.

The result is a new duality relating the 
two-dimensional DG model with arbitrary potential $v$ to 
another such model but with a different potential $\tv$ 
and, again, with {inverse} temperature {$\tbeta=\pi^2/\beta$}.  
More precisely,
the partition function on the torus  is shown to obey the duality
\beq
Z_{N,M}\left[\beta v\right] = \frac{1}{\beta^{(MN-1)/2}}  c_{N,M}[v]\,\,Z_{N,M}\left[\frac{\pi^2}{\beta}\tv\right],
\label{DualityZtorus}
\eeq
where the constant $c_{N,M}[v]$ is a functional of $v$,  and where the
relation between the potentials $v(\vecr)$ and $\tv(\vecr)$ is given in section
\ref{sec:DualityOnTorus} in terms of their Fourier transforms.
Again, there is a self-dual potential
$\vstar$ and a candidate critical temperature.

This paper is organized as follows.
In section \ref{sec:DefinitionDGmodels} 
we establish our notation for the 1D and 2D Discrete Gaussian models, and we recall the results about the  duality on the ring.
In section \ref{sec:Mapping},  by using the Chinese remainder theorem
we introduce and discuss the mapping between a one-dimensional and a two-dimensional
lattice  and the corresponding transformation of periodic functions.
In section \ref{sec:newDuality} we show how for the two-dimensional DG Hamiltonian
this mapping leads to a duality relation.
In section \ref{sec:Selfduality} we consider the special case of the self-dual
potential $\vstar$.
In section \ref{sec:nnbpotential}
we consider  the well-known 2D DG Hamiltonian with isotropic
nearest-neighbor interaction. 
In section \ref{sec:Conclusion} we point out the main features of the new duality.

\section{Discrete Gaussian models}
\label{sec:DefinitionDGmodels}

In this section we establish some notation and review some
results on the one-dimensional DG model that will be fundamental in the
sections hereafter.
The length of the ring will be denoted by $\calN$, a coordinate difference 
by $R$,  and the potential by $V(R)$.

\subsection{DG model on a ring}
\label{sec:DG1D}

For a  ring of length $\calN$ we shall write ${\calH}_1$
for the DG Hamiltonian  (\ref{HDGd}).
The lattice site $\veci$ becomes a scalar $i$
that may take the values $i=0,1,\ldots,\calN-1$.
In a slightly more formal notation we then have\\
\beq
{\calH}_1 = \frac{1}{2} \sum_{i\in \mathbb{Z}_\calN} 
\sum_{i^\prime\in \mathbb{Z}_\calN} 
V(i-i^\prime) (h_i-h_{i^\prime})^2,
\label{HDG1bis}
\eeq
where $i\in\mathbb{Z}_\calN$ is the equivalence class of
all integers equal to $i$ up to a multiple of $\calN$.
Symmetry of the interaction under parity transformation
is expressed as
\beq
V(R) = V(-R).
\label{symm11}
\eeq
Since the labels $i$ and $i+\calN$ refer to the same site,
the potential $V(R)$ must be  $\calN$-periodic, 
\beq
V(R) = V(R+\calN).
\label{symm12}
\eeq
The two equations (\ref{symm11}) and (\ref{symm12}) together imply the reflection symmetry
\beq
V(R) = V(\calN-R).
\label{symm13}
\eeq

We define the partition function with the global gauge 
mentioned in the introduction, namely
\beq
Z_\calN\left[\beta V\right]=\sideset{}{'}{\sum}_{\{h_i|i\neq 0\}}
\exp[{-\beta{\calHDG}_1}],
\label{Z1bis}
\eeq
where the prime indicates the constraint $h_0=0$. This restriction implies 
that the mean height at any site $i$ vanishes at any temperature.

We observe that Hamiltonian (\ref{HDG1bis})
is independent of the value of $V(0)$.
In the Fourier transforms below we shall consider that
$V(0)$ has been assigned an arbitrary value,
knowing that the results cannot depend on it.

Fourier transformed variables are defined as
\beq
\hh_K=\frac{1}{\sqrt{\calN}}\sum_{j\in \mathbb{Z}_\calN}\ee^{-\ii K  j}\, h_j\,,
\label{FTh1}
\eeq
and the Fourier transformed potential is
\beq
\hV(K) = \sum_{R\in \mathbb{Z}_\calN} \ee^{-\ii KR} \, V(R), 
\label{FTv1}
\eeq
where $K=2\pi p/\calN$ with $p\in\mathbb{Z}_\calN$. Then 
Hamiltonian (\ref{HDG1bis}) takes the form
\beq
\calHDG_1[V] =  \sum_{K\neq 0} W(K) \hh_K\hh_{-K}
\label{HDG1FT}
\eeq
in which 
\beq
W(K) \equiv \sum_{R=1}^{\calN-1} \left[1-\cos(KR)\right] V(R)= \hV(0)-\hV(K),
\quad K\neq 0.
\label{FTw1}
\eeq
The last equality in (\ref{FTw1}) comes from the symmetry (\ref{symm13}).
Equation (\ref{HDG1FT}) shows, incidentally, that in order
for the partition function 
(\ref{Z1bis}) to exist we must have that $W(K)>0$ for all $K\neq 0$;
we impose this condition throughout the remainder of this paper.

Equation (\ref{FTv1}) is the usual Fourier transform in the space of $\calN$
points with  inverse
\beq
V(R) = \frac{1}{\calN} \sum_K \ee^{\ii KR}\, \hV(K), \qquad R=0,1,\ldots,\calN-1.
\label{invFTv1}
\eeq
Equation (\ref{FTw1}) transforms only the subset 
$\{V(1),V(2),\ldots,V(\calN-1)\}$ which
excludes the nonphysical variable $V(0)$, and its inverse is
\beq
V(R) = -\frac{1}{\calN}\sum_{K\neq 0} \ee^{\ii KR}\, W(K), \qquad R=1,2,\ldots \calN-1.
\label{invFTw1}
\eeq

The symmetries (\ref{symm11})-(\ref{symm13}) lead for $\hV(K)$
to the corresponding symmetries 
\beq
\hV(-K) = \hV(K), \quad \hV(K+2\pi) = \hV(K),  \quad \hV(2\pi-K) = \hV(K),
\label{symmhv}
\eeq
of which only two are independent.
Clearly $W(K)$ satisfies the same symmetries as $\hV(K)$.

\subsection{Duality in one dimension}
\label{sec:dual1D}

It was shown in reference \cite{KjaerHilhorst82} 
that the one-dimensional DG model with arbitrary potential $V(R)$
obeying the symmetries (\ref{symm11})-(\ref{symm13})
is dual to a similar one-dimensional DG model with a potential
$\tV(R)$. In particular,
the partition functions of the two models are related by
\beq
Z_\calN\left[\beta V\right] = \frac{1}{\beta^{(\calN-1)/2}} C_{\calN}[V]\,\, Z_{\calN}\left[\tbeta\tV\right],
\label{DualityZring}
\eeq
where $\tbeta=\pi^2/\beta$ and
\beq
C_{\calN}[V]=\sqrt{\calN}\sqrt{ \prod_{K\neq 0} \frac{\pi }{W(K)}}
\label{DefConstant}
\eeq
with $K=2\pi p/\calN,\, p\in \mathbb{Z}_{\calN}$, and $W(K)$ defined  in
(\ref{FTw1}).\footnote{The partition function of the dual model depends  on
  $\tbeta\tV$ and in reference \cite{KjaerHilhorst82} the normalizations of the potential $\tV$ and the inverse temperature $\tbeta$ are such that $\tbeta=1/\beta$.}
The relation between $V(R)$ and $\tV(R)$ 
takes its simplest form in terms of $W(K)$ and $\tW(K)$, namely
\beq
\tW(K) = \frac{4\sin^2(K/2)}{W(K)}, \qquad K\neq 0.
\label{dualityTFW}
\eeq
When expressed in terms of $\hV(K)$ and $\htV(K)$ this relation becomes
\beq
\htV(0)-\htV(K) = \frac{ 2[1-\cos K]}{\hV(0)-\hV(K)}, 
\qquad
K\neq 0,
\label{dualityTFV}
\eeq
and leaves $\htV(0)$ undefined.
The real-space expression $\tV(R)$ of the dual potential
may be obtained by inverse Fourier
transformation of (\ref{dualityTFW}) according to (\ref{invFTw1})  with the result
\bea
\tV(R) &=& -\frac{1}{\calN}\sum_{K\neq 0} \ee^{\ii KR}\, 
\frac{4\sin^2(K/2)}{W(K)}
\nonumber\\[2mm]
&=& - \frac{1}{\calN} \sum_{K\neq0} \ee^{\ii KR}\, \frac{ 2[1-\cos K]}{\hV(0)-\hV(K)}, \qquad R=1,\ldots,\calN-1.
\label{invFTtV}
\eea
This equation leaves $\tV(0)$ undefined.
Furthermore, neither $W(K)$ nor $\tW(K)$
appears in the transformations with argument $K=0$.

We also notice that, according to (\ref{symmhv})  and (\ref{dualityTFV}),  $\htV(K)$ obeys the same reflection symmetry as $\hV(K)$, namely 
\beq
\htV(2\pi-K) = \htV(K).
\label{symFTDualV}
\eeq
As a consequence $\tV(R)$ obeys the same reflection symmetry as $V(R)$,
\beq
\tV(R)=\tV(\calN-R).
\label{symDualV}
\eeq

\begin{figure}[ht]
\begin{center}
\scalebox{.55}
{\includegraphics{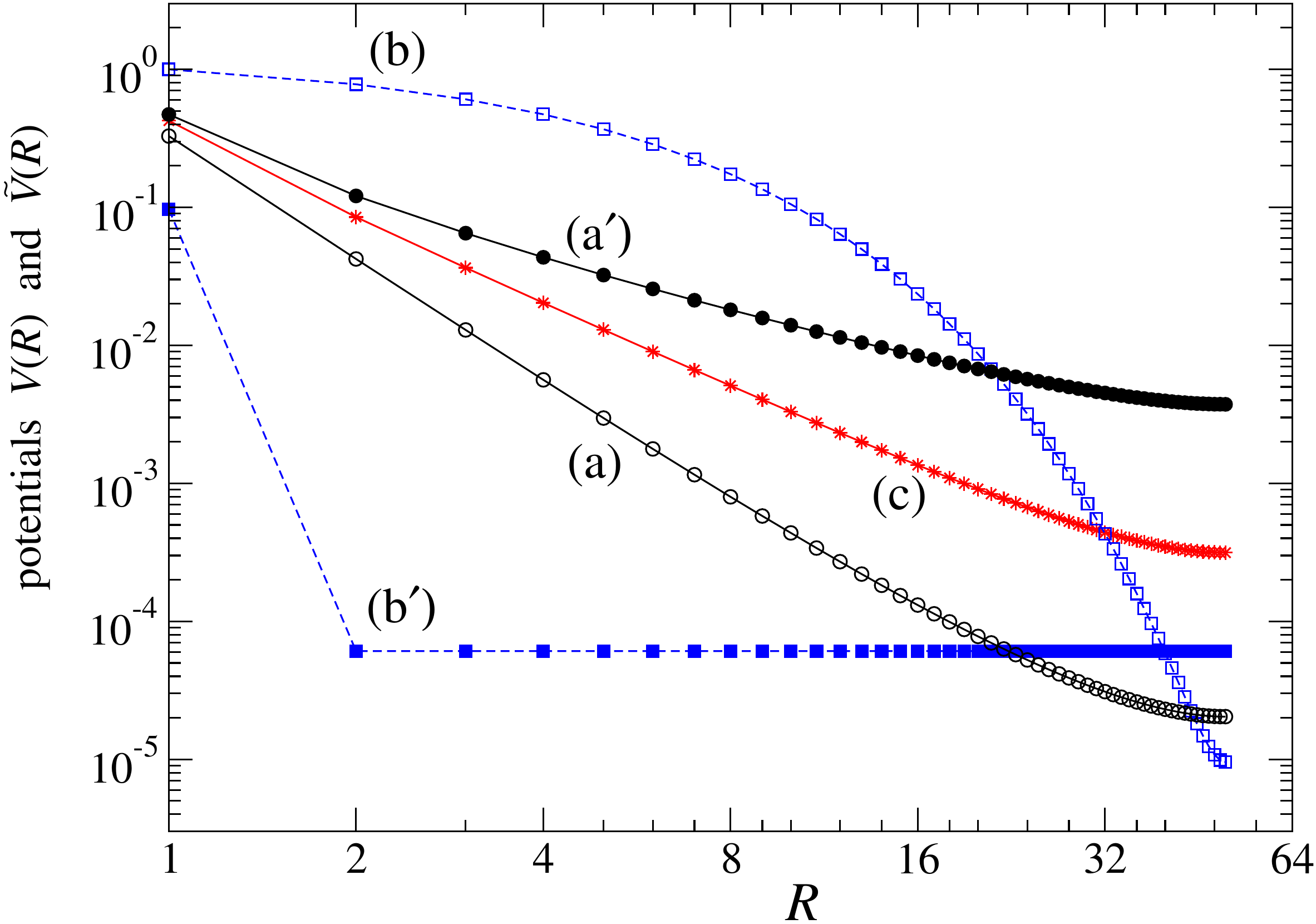}}
\end{center}
\caption{\small 
Three examples of a potential $V(R)$ and its dual $\tV(R)$ on a ring of
$N=100$ sites.
(a) Open black circles: The power law potential $V(R)=\pi^{-1}N^3/[R(N-R)]^3$,
appropriately symmetrized 
to satisfy equation (\ref{symm13}); for $N\to\infty$ it tends 
to $V(R)=1/(\pi R^3)$.
(a$^\prime$): Its dual (filled black circles);
for $N\to\infty$ it tends to a potential 
that decays with distance as $\sim1/R$. 
(b) Open blue squares: The exponential potential $V_{\rm exp}(R)$ 
of equation (\ref{dVexp}) for $\alpha=0.25$. 
(b$^\prime$): Its dual (filled blue squares),
equations (\ref{dtVexp}) and (\ref{xAB}).
(c) Red stars:
The self-dual potential $V(R) = \tV(R)=V^\star(R)$, equation (\ref{VstarR}).
}
\label{fig_one}
\end{figure}

In figure \ref{fig_one} we have represented two examples of a potential $V(R)$
and its dual. In general, the relation between $V(R)$ and $\tV(R)$ cannot
be made more explicit than equation (\ref{dualityTFV}) or equivalently 
(\ref{invFTtV}). Among the exceptions is the exponentially decaying
potential, appropriately symmetrized to satisfy equation (\ref{symm13}),
\beq
V_{\rm exp}(R) = \frac {\ee^{-\alpha R}+\ee^{-\alpha (N-R)}} {\ee^{-\alpha} + \ee^{-\alpha(N-1)}},
\label{dVexp}
\eeq
whose dual is 
\beq
\tV_{\rm exp}(R) = A(\delta_{R,1}+\delta_{R,N-1})+N^{-1}B
\label{dtVexp}
\eeq
with
\beq
A=\frac{ \cosh\frac{\alpha(N-2)}{2} \sinh\frac{\alpha}{2} }
       { \sinh\frac{\alpha N}{2}     \cosh\frac{\alpha}{2} }\,, 
\qquad B=4A\sinh^2\frac{\alpha}{2}\,. 
\label{xAB}
\eeq

For future use we introduce an auxiliary potential $U(R)$, 
\beq
U(R)=-\frac{1}{\calN} \sum_{K\neq 0}
\frac{1-\cos(KR)}{\hV(0)-\hV(K)}, \qquad R=0,1,2,\ldots,\calN,
\label{UDef}
\eeq
in which $\hV(0)-\hV(K)>0$.
It is easily checked that $\tV(R)$ may be derived from $U(R)$ by
\beq
\tV(R) =U(R+1)+U(R-1)-2U(R), \qquad R=1,2,\ldots,\calN-1,
\label{tVUquad}
\eeq
a relation that appears in reference \cite{KjaerHilhorst82}  (but with another
normalization). 
This $U(R)$ arises  through the well-known correspondence between a  DG model  and a lattice  model in which the Hamiltonian reads $ \frac{1}{2} \sum_{i\in \mathbb{Z}_\calN} 
\sum_{i^\prime\in \mathbb{Z}_\calN} 
U(i-i^\prime) q_iq_{i^\prime}$  and  the   configurations of integer  $q_i$'s
obey the  neutrality constraint $\sum_iq_i=0$. Therefore the integer-valued
$q_i$ are called ``charges''  and their interaction potential $U$ 
the ``charge potential.''
Consequently equation (\ref{tVUquad})
shows that $\tV(R)$ 
is the potential created by a quadru\-pole of charges $1$ ,$-2$, and
$1$ located on the sites $R=-1$, $R=0$, and $R=1$, respectively. 
We shall therefore sometimes refer to
$\tV(R)$ as the ``quadru\-polar interaction.''

\clearpage
\subsection{DG model on a torus}
\label{sec:DG2D}

In the special case of an $N\times M$
 lattice with toroidal boundary conditions
we shall write the DG Hamiltonian (\ref{HDGd}) as ${\calH}_2$.
Sites will be labeled by $\veci=(i,j)$, where $i=0,1,\ldots,N-1$ and 
$j=0,1,\ldots,M-1$.
The Hamiltonian (\ref{HDGd}) then becomes
\beq
 {\calH}_2 = \frac{1}{2} \sum_{(i,j)\in \mathbb{Z}_N\times \mathbb{Z}_M} \sum_{(i^\prime,j^\prime)\in \mathbb{Z}_N\times \mathbb{Z}_M}
v(i-i^\prime, j-j^\prime) (h_{i,j}-h_{i^\prime,j^\prime})^2.
\label{HDG2}
\eeq
Parity symmetry  is now expressed as
\beq
v(r,s) = v(-r,-s).
\label{symm21}   
\eeq
Since the labels $(i,j)$, $(i+N,j)$, and $(i,j+M)$ refer to the same site,  
the potential $v$ must have the periodicity properties
\beq
v(r+N,s) = v(r,s), \qquad v(r,s+M) = v(r,s).
\label{symm22} 
\eeq
As a consequence of (\ref{symm21}) and (\ref{symm22}) 
we have the reflection symmetry
\beq
v(N-r,M-s) = v(r,s).
\label{symm23}
\eeq
For $M=1$ this system reduces to the ring model described above.

Again, the Hamiltonian (\ref{HDG2}) is independent of the value of 
the interaction constant $v(0,0)$.
The partition function is defined as in (\ref{Z1bis}),
\beq
Z_{N,M}[\beta v]=
\sideset{}{'}{\sum}_{\{h_{i,j}\, \vert (i,j)\neq (0,0)\}}\exp[{-\beta{\calHDG}_2}],
\label{Z2}
\eeq
where the prime denotes the gauge condition $h_{0,0}=0$.

We shall consider in this work only functions on the ring and on the torus
that have the symmetry properties  (\ref{symm11})-(\ref{symm13}) and 
(\ref{symm21})-(\ref{symm23}), respectively.

\section{Mapping between a torus and a ring}
\label{sec:Mapping}

In this section we show how, under the condition that $N$ and $M$ are coprime,  the Chinese remainder theorem allows us to
introduce a mapping  between the ring $\mathbb{Z}_{NM}$ and  the torus  $\mathbb{Z}_N\times \mathbb{Z}_M$  for  both coordinates and   periodic functions. 
For the Chinese remainder theorem at an elementary level see reference
\cite{Rosen2010} and for more advanced topics see reference \cite{Lang2002}.

\subsection{Mapping for spatial coordinates}

\subsubsection{Chinese remainder theorem}

For any bijection of the sites $(r,s)$ of the torus $\mathbb{Z}_N\times
\mathbb{Z}_M$ onto the integers $R$ of the ring $\mathbb{Z}_{NM}$
the Hamiltonian (\ref{HDG2}) becomes formally a one-dimensional Hamiltonian. 
We wish, however, to apply a bijection that
preserves the group law ({\it i.e.,} translation and inversion). 
The Chinese remainder theorem  
provides such a bijection
at the condition that $N$ and $M$ be coprime, that is, that their only positive common divisor be unity.
We shall henceforth take $M$ and $N$ such that this condition is met.
 
 In the case of two integers  $N,M>1$ the Chinese remainder theorem may be stated as follows.
For any given pair of integers $(r,s)$  the set of equations with unknown $R$,
\beq
r = R  \pmod{N},  \qquad s = R \pmod{M},
\label{forward}
\eeq 
where $x = y  \pmod{N}$ means that $x$ and $y$ differ  by a multiple of $N$,
has a solution given by
\begin{subequations} \label{backward}
\begin{equation} \label{backward1}
R= aNs +bMr \pmod{NM},
\end{equation}
in which the pair of integer {\it B\'ezout coefficients\,} $(a,b)$
is, in turn, a solution of
\begin{equation} \label{backward2}
aN+bM =1.
\end{equation}
\end{subequations}
Bézout's theorem guarantees that there exists 
a pair $(a,b)$ satisfying (\ref{backward2})
which may be found by the so-called extended Euclidean algorithm. The linear combination in (\ref{backward1}) is readily shown to satisfy the set of equations (\ref{forward}) as follows. By construction $aN s +bM r= b M r \pmod{N}$; then according to the identity (\ref{backward2}),   $bM r$ can be rewritten as $r-aN r$ and  $r-aN r=r\pmod{N}$. As a result $aN s +bM r=  r \pmod{N}$. A similar argument leads to 
 $aN s +bM r=  s \pmod{M}$.

We notice that from the definition (\ref{backward2})
 of the Bézout coefficients  it immediately follows that another pair of the form $(a+cM,b-cN)$, where $c$ is any integer, is also a solution. 
We may make the solution of \eqref{backward2} unique by imposing, for example, that $0<a<M$ and $-N<b<0$,
or, alternatively, that $-M<{a}<0$ and $0<{b}<N$. With the constraint $0<a<M$ and $-N<b<0$, we give the  solutions  for various special cases  of $N$ and $M$.  For $N>1$ and $M=2$, the solution is
$a=1$ and $b=(1-N)/2$;
for $N>2$ and $M=N-1$  one gets $a=1$ and $b=-1$;
and for $M>1$ and $N=qM+1$,
where $q=2,3,\ldots$, one finds $a=1$ and $b=-q$.

\clearpage
\subsubsection{Geometrical interpretation} 
\label{sec:UneBijection}

\begin{figure}[ht]
\begin{center}
\scalebox{.45}
{\includegraphics{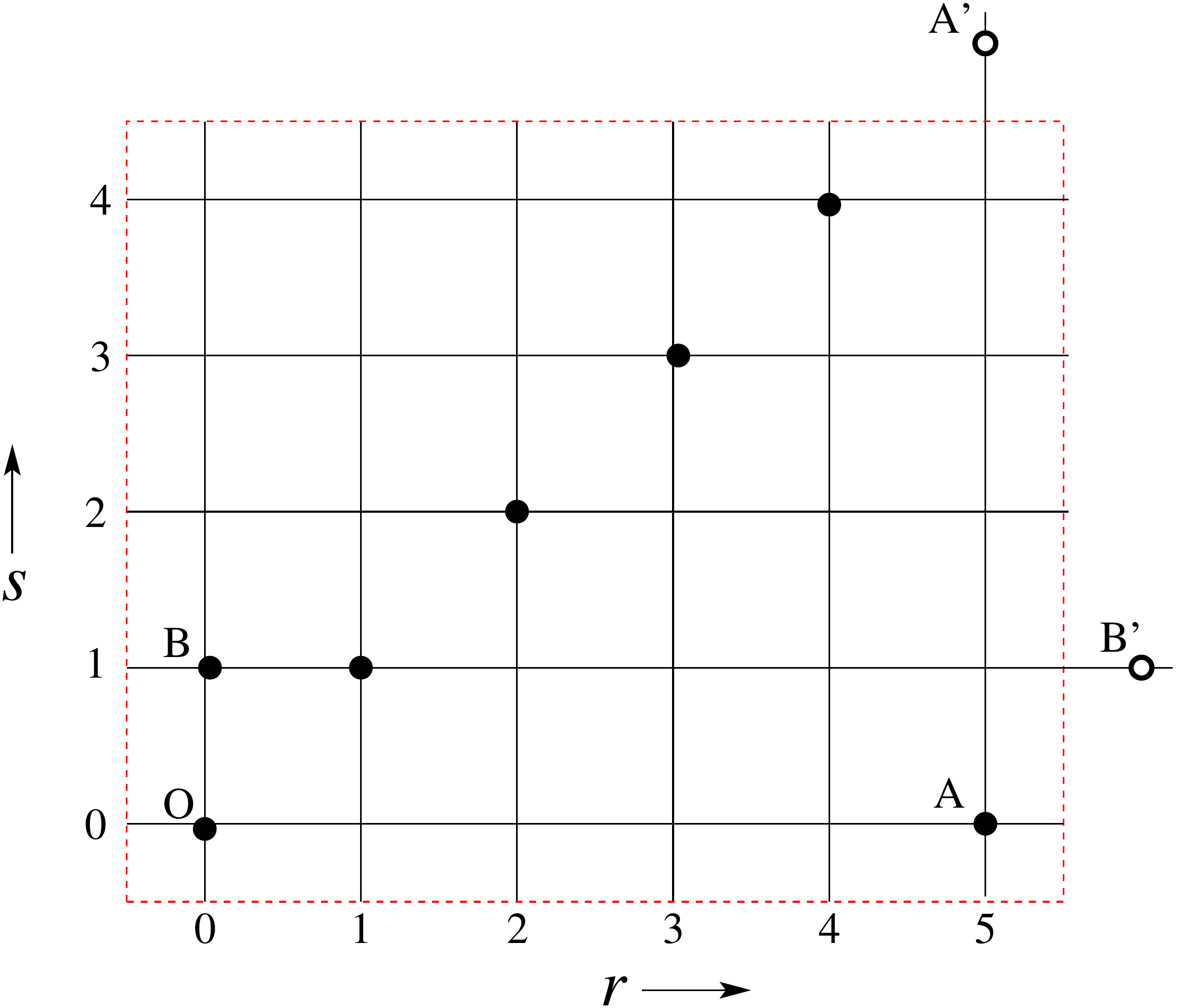}}
\end{center}
\caption{\small 
The sites of an $N\times M$ toroidal lattice with $N=6$ and $M=5$.
The fundamental domain is the region inside the dashed red rectangle,
which must be considered as periodically repeated; in particular, $A$
and $A^\prime$ are identical sites, and so are $B$ and $B^\prime$.
The path described in the text starts at the origin $O$ and moves at each step
diagonally in the direction $(1,1)$, as indicated by the black dots.
When it steps to site $A^\prime$,
it leaves the fundamental domain but in fact arrives at $A$. Its subsequent
step in the $(1,1)$ direction then takes it to $B^\prime$, 
but in fact it arrives at $B$. Upon continuing it will visit
all lattice sites until at its $NM$th step it returns to $O$. 
}
\label{fig_two}
\end{figure}

The Chinese remainder theorem may be interpreted as a helicoidal mapping
of a one-dimensional path around the $N\times M$ torus in the following
way. We refer to figure \ref{fig_two}.
We let $R$ take the successive values $R=0,1,2,\ldots,NM$
and consider the path traced out on the torus by the pair $(r,s)$ 
parametrized by $R$ according to equation (\ref{forward}).
At $R=0$ the path starts in the origin 
$(r,s)=(0,0)$, and 
as long as $R < \min(N,M)$ we have 
$(r,s)=(R,R)$, that is, the path follows the main diagonal,
undergoing at each step an increment $(1,1)$.
For larger $R$ the path continues to undergo increments $(1,1)$,
but the $N,M$-periodicity of the lattice has to be taken into account.
This leads to a path that winds
around the torus until it returns to the origin.
The condition that $N$ and $M$ be coprime guarantees
that this return will occur only after the path has visited all
sites of the torus. 

\subsubsection{Another mapping and corresponding helicoidal winding}
\label{sec:AutreBijection}

In section (\ref{sec:MappingPeriods}) we shall show that  the functions $F(R)\equiv f(r,s)$ defined  with the mapping (\ref{backward1}) are periodic with period $MN$.
We notice that we could have chosen another helix  which winds around the torus while also preserving the periodicity of the lattice, so that   the corresponding mapping  $(r,s)\mapsto R'$ leads to the same periodicity $NM$ for functions $F'(R')\equiv f(r,s)$.

For instance another proper  helix is built by again  mapping   the origin $(r,s)=(0,0)$ onto the integer $R=0$ and then by incrementing the position on the $N\times M$ lattice by steps of $(-1,1)$ while at the same time increasing $R$ by one unit. After $R'$ steps on the helix the corresponding coordinates on the torus are
\beq
r = -R'  \pmod{N}, \qquad s = R' \pmod{M}.
\label{Helix2}
\eeq
A simple argument similar to that presented for the derivation of the Chinese theorem (\ref{backward}) shows that the linear combination 
\beq
R'=a  N s  - bM r
\label{Mapping2}
\eeq
is a solution of  (\ref{Helix2}), because $a$ and $b$  are the solutions of 
(\ref{backward2}). We shall see in subsection (\ref{sec:DualityVersusMapping}) how the results of interest in the present paper depend on the choice of one among the two   mappings $R=aNs+bMr$ or $R'=a  N s  - bM r$.

\subsection{Mapping for periodic functions}
\label{sec:MappingPeriods}

\subsubsection{Periodicity  on the torus and on the ring}

Let $f(r,s)$ be a given biperiodic 
function on the $N\times M$ torus obeying the symmetry properties
(\ref{symm21})-(\ref{symm23}). 
\beq
f(r+N,s)=f(r,s+M)=f(r,s)
\label{periodicityf1}
\eeq
and
\beq
f(N-r,M-s)=f(r,s).
\label{periodicityf2}
\eeq
We define a corresponding function $F(R)$ on the ring of length $NM$ by
\beq
F(R)=f(r,s)\quad\textrm{with}\quad R= aN s+bM r  \pmod{NM}.
\label{Fring}
\eeq
We shall show that the symmetries of $f$ imply those of $F$ and that the reciprocal is also true.

The pairs 
$(r+N ,s)$ and $(r,s+M)$ are associated with  $R+bNM$
and $R+aNM$, respectively. 
The periodicity properties  (\ref{periodicityf1}) then lead to \break
$
F(R+bMN)= F(R+aMN)=F(R).
$
It follows that 
$F(R)=\break F(R+N(aNM)+ M(bNM))$, 
whence, with the use of \eqref{backward2}, we find 
that
\beq
F(R+NM)=F(R).
\label{PeriodicityF}
\eeq
Similarly, by virtue of 
\eqref{backward2}, the reflected  point $(N-r,M-s)$ is associated with  
$aN(M- s) + bM(N-r)=NM-R$,  and the symmetry property (\ref{periodicityf2}) leads to
\beq
F(NM-R)=F(R).
\label{symmF}
\eeq
As a consequence we also have the third symmetry, $F(-R)=F(R)$;
that is, $F$ obeys the symmetries (\ref{symm21})-(\ref{symm23}) for a ring of
$NM$ sites.
We point out  that similar arguments show that the periodicity 
(\ref{PeriodicityF}) and the reflection symmetry (\ref{symmF}) are also valid for the second mapping (\ref{Mapping2}).

Let conversely $F(R)$ be given and satisfy (\ref{PeriodicityF}) and 
(\ref{symmF});  then \break
$f(r+N,s)=f(r,s+M)=f(r,s)$ and  $f(N-r,M-s)=f(r,s)$ . Indeed if \break $F(R+NM)=F(R)$ then $F(R+aNM)=F(R)$, and $R+aNM$ corresponds to $(r,s+M)$. Similarly if $F(NM-R)=F(R)$ then $F((a+b)NM-R)=F(R)$, and $(a+b)NM-R$ corresponds to $(N-r,M-s)$.

\subsubsection{Mapping  Fourier transforms from the torus to the  ring}

The Fourier transform of an $NM$-periodic function $F(R)$ on the
ring  of length $NM$ is
\beq
\widehat{F}(K)=\sum_{R\in \mathbb{Z}_{NM}} \ee^{-\ii K R} \, F(R)
\label{FTring}
\eeq
with the wavenumbers
\beq
K=\frac{2\pi}{NM}p,
\label{rangeK}
\eeq
where $p \in \mathbb{Z}_{NM}$.
Similarly, the Fourier transform of an $N,M$-biperiodic function $f(r,s)$ on
the torus is
\beq
\widehat{f}(\kun,\kdeux) = 
\sum_{r\in \mathbb{Z}_N}\sum_{s\in \mathbb{Z}_M} \ee^{-\ii(\kun r +\kdeux s)} \, f(r,s)
\label{FTtorus}
\eeq
with the wavenumbers 
\beq 
\kun=\frac{2\pi}{N} n, \qquad \kdeux=\frac{2\pi}{M} m,
\label{rangekxky}
\eeq
where $n\in\mathbb{Z}_N$ and $m\in\mathbb{Z}_M$.
We now investigate the relation that results
between these two Fourier transforms in case $F(R)=f(r,s)$.

The Chinese remainder theorem allows us 
to establish a bijection between the index
$p$  of the wavenumber on the ring and the 
index pair $(n,m)$ on the torus,
\beq
p= a N m + bM  n\pmod{NM},
\label{defp}
\eeq
which is analogous to $R= aN s + bM r \pmod{NM}$. 
Hence the wavenumber on the torus may be expressed in terms of
those on the ring as 
\beq
K= b\kun+ a\kdeux.
\label{decompK}
\eeq
After some rewriting and use of identity (\ref{backward2}) we find 
\beq
\ee^{-\ii K R}=\ee^{-\ii ( b  \kun r  + a \kdeux s )}.
\label{expIKR}
\eeq
When we substitute (\ref{decompK}) in (\ref{FTring})
and identify $F(R)=f(r,s)$, we obtain
\beq
\widehat{F}(K) = \sum_{r\in \mathbb{Z}_N} 
\sum_{s\in \mathbb{Z}_M} \ee^{-\ii( b  \kun r  + a \kdeux s )} f(r,s)
=\widehat{f}(b \kun, a\kdeux).
\label{FTFKfromFTfmunu}
\eeq
The Fourier transform $\widehat{F}(K)$ on the ring proves to coincide with  a scaled Fourier transform on the torus.
Relation (\ref{FTFKfromFTfmunu}) allows to determine the Fourier transform $\widehat{F}(K)$ on the ring when the Fourier transform $\widehat{f}(\kun, \kdeux)$ on the torus is given.

\subsubsection{Mapping   Fourier transforms from the  ring to the torus}

We shall now see how to determine the Fourier transform $\widehat{f}(\kun, \kdeux)$ on the torus  when the Fourier transform $\widehat{F}(K)$ on the ring  is given.
We first write  $f(r,s)=F(R)=\frac{1}{MN}\sum_{K} \ee^{\ii KR}
\widehat{F}(K)$, and from (\ref{decompK}) and (\ref{expIKR}) we get
\beq
f(r,s)=\frac{1}{MN} \sum_{(\kun,\kdeux)}\ee^{\ii ( b  \kun r  + a \kdeux s )} \widehat{F}(b\kun+a \kdeux).
\label{fVersusFTFK}
\eeq
where $(\kun,\kdeux)$ is related to a pair of integers $(n,m)$ through
(\ref{rangekxky}).

We then notice that, according to (\ref{backward2}), the coefficient $b$ is coprime with $N$ (because if  $b$ and $N$ had a common divisor different from $1$ or $-1$ then  $aN+bM$ could  not be equal to $1$). By virtue of Gauss's lemma, the fact that   there exists no common divisor of  $b$ and $N$
entails that if $b(n-n')$ is a multiple of $N$, then $n-n'$ is also a multiple of $N$. Equivalently $n\neq n'\pmod{N}$ implies $bn\neq bn'\pmod{N}$ and  $n\mapsto bn$ is a one-to-one correspondence  from $ \mathbb{Z}_N$ to $ \mathbb{Z}_N$.
Similarly, according to (\ref{backward2}),  the coefficient $a$ is coprime with $M$ and $m\mapsto am$ is a one-to-one correspondence  from $ \mathbb{Z}_M$ to $ \mathbb{Z}_M$.
As a result, if the function $A(n,m)$ is $N,M$-periodic, 
then 
\beq
 \sum_{(n,m) \in \mathbb{Z}_N\times \mathbb{Z}_M} A(bn,am)
  =\sum_{(n,m) \in \mathbb{Z}_N\times \mathbb{Z}_M} A(n,m).
  \label{SumInvariance}
\eeq
 Hence  the sum in   (\ref{fVersusFTFK}) can be rewritten without the coefficients $a$ and $b$, and eventually
\beq
f(r,s)= \frac{1}{MN} \sum_{(\kun,\kdeux)}\ee^{\ii (\kun r +\kdeux s )} \widehat{F}(\kun+ \kdeux).
\label{fVersusFTFKBis}
\eeq
Upon Fourier transforming both members of this equation 
we find
\beq
\widehat{f}(\kun,\kdeux)= \widehat{F}(\kun+\kdeux),
\label{FTFKtoFTfmunu}
\eeq
which is the desired relation that yields $\widehat{f}(\kun,\kdeux)$ when 
$\widehat{F}(K)$ is given.

\section{New duality for two-dimensional DG models}
\label{sec:newDuality}

\subsection{Mapping between torus and ring Hamiltonians}
\label{sec:maptorusring}

In the preceding section we have  defined a mapping between the sites of the torus and those of the
ring,  and an identification of functions defined on the torus with functions
defined on the ring. Now we consider how a Hamiltonian given on the torus
transforms into one defined on the ring.

Let the Hamiltonian ${\calH}_2$ of equation (\ref{HDG2}) be given.
A mapping of this Hamiltonian, defined on the torus $\mathbb{Z}_N\times
\mathbb{Z}_M$, onto a Hamiltonian on the ring $\mathbb{Z}_{NM}$ is constructed
as follows. 
We relabel the height variables $h_{r,s}$  according to 
\beq
h_R=h_{r,s}\,,
\eeq
where $R$ is given by (\ref{backward}) 
and we define the potential $V(R)$ by
\beq
V(R)=v(r,s).
\eeq
According to section (\ref{sec:MappingPeriods}) the periodicity properties (\ref{symm22}) and the reflection symmetry (\ref{symm23})  of $v(r,s)$ 
imply that
\beq
V(R)=V(R+MN)
\eeq
and
\beq
V(R) = V(MN-R).
\label{symmV}
\eeq
When we express the two-dimensional DG Hamiltonian  ${\calH}_2$ defined in (\ref{HDG2}) in terms of the new quantities $V(R)$ and
$h_R$, we find that 
${\calH}_2$ becomes a one-dimensional DG Hamiltonian of type (\ref{HDG1bis}),
\beq
\calHDG_1 [V]= \frac{1}{2} \sum_{i\in\mathbb{Z}_{MN}}\sum_{R\in\mathbb{Z}_{MN}}V(R)
(h_{i}-h_{i+R})^2,
\label{HDG1mapped}
\eeq 
and the partition functions of the two models are identical,
\beq
Z_{N,M}\left[\beta v\right] = Z_{NM}\left[\beta V\right].
\label{RelZtorusZring}
\eeq
Hence we have identified the partition function on the torus with a partition function on the ring.

\subsection{Duality on the torus}
\label{sec:DualityOnTorus}

Relation (\ref{RelZtorusZring}) embodies the mapping of a given two-dimensional system with
potential $v$ onto a one-dimensional one with related potential $V$.
We may now apply, without recalling all the intermediary steps, 
the mechanism of section \ref{sec:dual1D}  whereby
$V(R)$ is related to a dual one-dimensional potential $\tV(R)$.
Subsequently we return to a dual two-dimensional potential $\tv(r,s)$
by means of the relation
\beq
\tv(r,s)=\tV(R),
\eeq
Because of (\ref{symDualV}) the dual potential also has the reflection property on the torus
\beq
\tv(r,s) = \tv(N-r,M-s).
\label{symrefltvMN}
\eeq
The corresponding DG partition function is given by the identity
(\ref{RelZtorusZring}),
\beq
Z_{N,M}\left[\beta \,\tv\right]=Z_{NM}\left[\beta \,\tV\right].
\label{RelZtorusZringDual}
\eeq
By combining the duality relation (\ref{DualityZring}) between the partition
functions on the ring with (\ref{RelZtorusZring}) and
(\ref{RelZtorusZringDual}) we obtain
the two-dimensional duality
\beq
Z_{N,M}\left[\beta \,v\right] = \frac{1}{\beta^{(MN-1)/2}} C_{NM}[V]\,\,Z_{N,M}\left[\frac{\pi^2}{\beta} \,\tv\right].
\label{DualityZtorus1}
\eeq

In the relation (\ref{DualityZtorus1}) the constant $C_{NM}[V]$, given in (\ref{DefConstant}),
is still a functional  of the intermediate one-dimensional potential $V$.
We re-express it as follows as
a functional $c_{N,M}[v]$ of $v$. Indeed, $W(K)=\hV(0)-\hV(K)$ and, according
to (\ref{FTFKfromFTfmunu}) a Fourier transform on the ring is equal to a
scaled Fourier transform on the torus. Hence we have
\bea
C_{NM}[V]&=&\sqrt{NM}\left[ \prod_{(n,m)\in  \mathbb{Z}_N\times \mathbb{Z}_M\backslash(0,0)}
\frac{\pi }{\hv(0,0)-\hv(2\pi bn/N, 2\pi am/M)}\right]^{1/2}
\nonumber\\[2mm]
&=& \sqrt{NM}\left[ \prod_{(\kun,\kdeux)\neq(0,0)}
\frac{\pi }{\hv(0,0)-\hv(\kun,\kdeux)}\right]^{1/2}
\nonumber\\[2mm]
&\equiv&c_{N,M}[v]
\label{ConstantTorus}
\eea
where to arrive at the second line we have used the property
(\ref{SumInvariance}), and we used the notation 
$\kun=2\pi n/N$ and $\kdeux=2\pi m/M$. 
Eventually the duality  relation (\ref{DualityZtorus1}) between partition functions on the torus reads in terms of functions defined on the torus 
\beq
Z_{N,M}\left[\beta v\right] = \frac{1}{\beta^{(NM-1)/2}}  c_{N,M}[v]\,\,Z_{N,M}\left[\frac{\pi^2}{\beta} \tv\right].
\label{DualityZtorusBis}
\eeq
This achieves the purpose of establishing a duality relation for
partition functions on the torus.

The relation between the given potential $v(r,s)$ and its dual $\tv(r,s)$
may be rendered more explicit.
As in the one-dimensional case, $\tv(r,s)$ may be re-expressed in terms of the Fourier transform of $v(r,s)$ as follows. According to (\ref{FTFKtoFTfmunu}) the Fourier transform  on the torus for $\tv(r,s)$ is given  in terms of   the Fourier transform  on the ring for $\tV(R)$ by
$\htv(\kun,\kdeux)=\htV(\kun+\kdeux)$, while the expression  for $\tV(K)$  in terms of $V(K)$ is given by (\ref{dualityTFV}). As a result
$\htv(\kun,\kdeux)-\htv(0,0)=2[1-\cos(\kun+\kdeux)]/[\hV(\kun+\kdeux)-\hV(0)]$. By using again relation (\ref{FTFKtoFTfmunu}) to go back from the ring to the torus,  namely $\hV(\kun+\kdeux)=\hv(\kun,\kdeux)$, we find that 
the Fourier transform of the dual potential  on the torus takes the simple
form 
\beq
\htv(\kun,\kdeux)-\htv(0,0)=-\frac{2[1-\cos(\kun+\kdeux)]}{\hv(0,0)-\hv(\kun,\kdeux)}.
\label{tvFT}
\eeq
Subsequently the expression of the dual potential $\tv(r,s)$ in terms of $v(r,s)$ is given by the inverse Fourier transform on the torus, 
for $(r,s)\neq(0,0)$
 \beq
\tv(r,s)=\ -\frac{1}{MN} 
\sum_{(\kun,\kdeux)\neq(0,0)}
\ee^{\ii [\kun r + \kdeux s]} \frac{ 2[1-\cos (\kun+\kdeux) ]} {\hv(0,0)-\hv(\kun,\kdeux)}
\label{tvInverseFTBis}
\eeq
with $\hv(0,0)-\hv(\kun,\kdeux)>0$ for $(\kun,\kdeux)\neq (0,0)$.
An explicit example of the duality embodied by equation (\ref{tvFT})
will be considered in section \ref{sec:nnbpotential}.

Finally, we may check that the square of the duality transformation is the
identity. Indeed, iteration of the duality relation (\ref{DualityZtorusBis})
leads to $Z_{N,M}\left[\beta v\right]=(1/\pi)^{NM-1}\,
c_{N,M}[v]c_{N,M}[\tv]\,Z_{N,M}\left[\beta \widetilde{\tv}\right]$, where
$\widetilde{\tv}=v$ according to (\ref{tvFT}), 
while the identity $\prod_{K\neq 0} 2 \vert \sin(K/2)\vert  = NM$ implies
that $c_{N,M}[v]c_{N,M}[\tv]$ $= c_{N,M}[V]c_{N,M}[\tV]$ $ = \pi^{NM-1}$.

We  notice that for a given mapping 
the  expressions for the constant $c_{N,M}$  and the dual potential are
independent of the Bézout coefficients $(a,b)$ according to  
(\ref{ConstantTorus})  and (\ref{tvInverseFTBis}).
As a result we could have chosen the pair of  Bézout coefficients $(\overline{a},\overline{b})$ such that $-M<\overline{a}<0$ and $0<\overline{b}<N$ 
with the mapping $\overline{R}=\overline{a} N s+\overline{b} Mr \pmod{NM}$ without changing the duality relation between the partition functions nor the expression of the dual potential $\tv$ in terms of the potential $v$.

\subsection{Dependence of the dual potential upon the choice of the mapping}
\label{sec:DualityVersusMapping}

As noticed above, the constant $c_{N,M}$ as well as  the relation between the  Fourier transforms of the dual potentials on the torus prove to be independent of $a$ and $b$  for a given mapping. However the dependence upon the choice of the mapping can be exemplified by the comparison of the two mappings presented in section (\ref{sec:UneBijection}) and (\ref{sec:AutreBijection}).

With the mapping $R=aNs+bMr$, the coordinates $R$  and $R\pm 1$  correspond to $(r,s)$ and $(r\pm 1, s\pm1)$, respectively. Then  the relation (\ref{tVUquad}) between $\tV(R)$ and $U(R)$ implies that  $\tv(r,s)=V(aNs+bMr)$ may be rewritten as 
\beq
\tv(r,s)=u(r+1,s+1) + u(r-1,s-1) -2 u(r,s),
\label{tvuquad}
\eeq
where, by using (\ref{UDef}),
\beq
u(r,s)=\frac{1}{MN} \sum_{(\kun,\kdeux)\neq (0,0)} 
\left[ \cos(\kun r +\kdeux s)-1\right]\frac{1}{\hv(0,0)-\hv(\kun,\kdeux)}.
\label{udef}
\eeq
 (The latter relation may also be directly derived from the inverse Fourier transform representation (\ref{tvInverseFTBis}) for $\tv(r,s)$,  as was done to derive (\ref{tVUquad})-(\ref{UDef}) from (\ref{invFTtV}).) 
  With the other mapping the coordinates $R'=aNs-bMr$ and  $R'\pm 1$
  correspond to $(r,s)$ and $(r\mp 1, s\pm1$), respectively. Then relation
  (\ref{tVUquad}) on the ring implies that  $\tv'(r,s)=V(aNs-bMr)$ and may be rewritten as  
\beq
\tv'(r,s)=u(r+1,s-1) + u(r-1,s+1) -2 u(r,s),
\label{tvuprimequad}
\eeq
with the same potential $u(r,s)$ as in relation (\ref{tvuquad}) for  the first mapping.

With the terminology introduced after (\ref{tVUquad}),
in the case of  the first mapping $\tv(r,s)$ appears as a  quadrupolar charge interaction, with  charges $(1,-2,1)$ aligned at points $(-1,-1)$, $(0,0)$, and $(1,1)$, respectively, along the direction of the first mapping helix. For the second  mapping $\tv'(r,s)$ still appears as a  quadrupolar charge interaction with  the same charge triplet, but the charges are  located at different points, namely $(-1,1)$, $(0,0)$, and $(1,-1)$, respectively,  along the direction of the second mapping helix at a given point.

The interaction $u$  is definitely independent of the mapping by virtue of (\ref{udef}). However the
above discussion  shows that   the dual potential $\tv$ depends on the mapping since it is a  
quadrupolar  interaction (involving  the  charge-charge interaction $u$) and  the locations of
the charges in the quadrupole depend on the mapping.

This investigation ultimately proves that the different options presented in section \ref{sec:Mapping} for the choice of the path in figure \ref{fig_one} amount to 
carrying out reflection symmetries with respect to the $r$ and/or $s$ axes.
They do not result in  any essentially new dualities. 

\section{Self-duality}
\label{sec:Selfduality}

\subsection{Self-dual potential and self-dual temperature}

As shown in reference \cite{KjaerHilhorst82}, the relation between the
potentials $V(R)$ and $\tV(R)$ on the ring, which is given by relation
(\ref{dualityTFV}) between their Fourier transforms,  leads to the existence
of a {\it self-dual potential\,} $\Vstar(R)$ such that for any $R\neq0$
\beq
\widetilde{\Vstar}(R)=\Vstar(R).
\label{DefVselfdual}
\eeq
Indeed, according to (\ref{dualityTFV}), if for $K\neq0$
\beq
\hVstar(K)-\hVstar(0)= - 2\vert\sin (K/2)\vert,
\label{hVstarK}
\eeq
then $\widehat{\widetilde{\Vstar}}(K)-\widehat{\widetilde{\Vstar}}(0)=\hVstar(K)-\hVstar(0)$, namely 
$\widetilde{W^\star}(K)= W^\star(K)=2\vert\sin (K/2)\vert$. The expression for
$\Vstar(R)$ when $R\neq0$ is obtained by inserting (\ref{hVstarK})
in  (\ref{invFTw1}).
The potential $\Vstar(R)$ is periodic in $R$ with period $MN$ and it may be written in various forms. For the following discussion we write
\beq
\Vstar(R) = \frac{\frac{1}{MN}\sin\frac{\pi}{MN}}
{\sin^2(\frac{\pi}{MN} R)-\sin^2(\frac{\pi}{2MN})}.
\label{VstarR}
\eeq

For the corresponding self-dual potential  on the torus,
$\vstar(r,s)=\Vstar(R)$  with $R=aN s+ bM r$. Moreover
$c_{N,M}[\vstar]=C_{NM}[\Vstar]=\pi^{(NM-1)/2}$ according to definition
(\ref{DefConstant}) and  the identity $\prod_{K\neq 0} 2 \vert \sin(K/2)\vert
= NM$. Therefore when $v=\vstar$ the duality relation (\ref{DualityZtorusBis}) for partition functions becomes  
\beq
Z_{N,M}\left[\beta\vstar\right] = \left(\frac{\betastar}{\beta}\right)^{(NM-1)/2} \,\,Z_{N,M}\left[\frac{{\betastar}^2}{\beta} \vstar\right]
\label{SelfDualityZtorus}
\eeq
with $\betastar=\pi$. This equation shows that there is a 
{\it self-dual (inverse) temperature\,} $\beta=\betastar=\pi$ at which 
(\ref{SelfDualityZtorus}) becomes a trivial identity.
In the next two sections we shall first investigate the self-dual potential 
$\vstar$ and then the height-height correlation function for this potential
when the system is at the dual temperature $\beta=\betastar$.

\subsection{Self-dual potential for large $N$}
\label{sec:selfdualThermoLimit}

We now investigate some of the properties of this two-dimensional self-dual potential.
We wish to consider its  limit  for a strip of
infinite length and finite width,  $N\to\infty$ with $M$ fixed, and for an infinite lattice, 
$N\to\infty$ and $M\to\infty$. By virtue of (\ref{VstarR}) the explicit
expression of  $\vstar(r,s)=\Vstar(R)$ is in fact a function of $R/(NM)$. 
In order to study the large-$N$ limit  of $\vstar(r,s)$  it is convenient to
make the change of  variables $(r,s)\mapsto (r,t)$
with
\beq
t=s-r,
\label{tsr}
\eeq
which, with the use of the identities (\ref{backward}), leads to rewriting  $R$ as $R=aNt +r$. Then $R/(NM)$ becomes
\beq
\frac{R}{MN}=\frac{at}{M}+\frac{r}{MN},
\label{xRtr}
\eeq
and, according to  (\ref{VstarR}),  the self-dual potential $\vstar(r,s)$  becomes the function
\beq 
\vstar(r,r+t) = \frac{\frac{1}{MN}\sin\frac{\pi}{MN}}
                {\sin^2\pi\big( \frac{at}{M} + \frac{r}{MN}\big) -   \sin^2\big(\frac{\pi}{2MN} \big)}.
\label{vstarsiBis}
\eeq

For coordinate differences $(r,r)$ we have that $t=0$ and  
\beq
\vstar(r,r)= \frac{\frac{1}{MN}\frac{\pi}{MN}}
                {\sin^2\pi\big( \frac{r}{MN}\big) - \sin^2\big( \frac{\pi}{2MN} \big)},
 \label{vstarrr}
\eeq
which depends only on $MN$. Therefore 
when $N$  goes to infinity, and whether or not $M$ remains finite, 
equation (\ref{vstarrr}) gives
\beq
\label{vrrNInfty}
\lim_{N\to\infty} \vstar(r,r) =  \frac{1}{\pi\left[r^2-\tfrac{1}{4}\right]}\,.
\eeq

For coordinate differences $(r,s)$ with $s\neq r$ we have to distinguish between $M$ remaining finite or tending 
to infinity, and we must know the Bézout coefficient $a$ as a function of $N$ and $M$. We 
shall choose to take
\beq
N = q M+1,
\label{ScalingLattice}
\eeq
with $q$ an arbitrary positive integer,
which ensures that $N$ and $M$ are coprime. In this case $a=1$ and $b=-q$. Then, by virtue of (\ref{vstarsiBis}),
$\vstar(r,r+t)$ becomes a function of $\frac{t}{M}+\frac{r}{MN}$.

In order to study  strips of finite width $M$ we consider the scaling
(\ref{ScalingLattice}) with $M$ fixed and $q\to\infty$, whence $N\to\infty$. 
For $M=2$ the torus is  the ladder lattice with each  interchain bond counting
twice, and for $M=4$ it is a 
beam with a square section.
Then for $r$ and $t\neq0$ fixed, $t/M$ remains finite while $r/(MN)$
vanishes. 
Upon inserting this limit behavior in equation (\ref{vstarsiBis})
and restoring the original coordinates $r$ and $s$ we find
\beq
\vstar(r,s) 
 \simeq \frac{\pi}{M^2 N^2 \sin^2\pi\big( \frac{r-s}{M} \big)}\,, 
  \qquad N\to\infty, \,\, M \mbox{ fixed}.
\label{vstarAymptoticrneqs}
\eeq

A two-dimensional infinite lattice is obtained when both $M$ and $N$ go to infinity with $q$ fixed.
For $r$ and $t$ fixed, $r/MN$ and $1/(2MN)$ vanish faster than $t/M$, and expression (\ref{vstarsiBis}) tends to the limit
\beq
 \vstar(r,s) \simeq \frac{1}{N^2\pi (r-s)^2}.
 \label{vstarAsOffDiagonal}
\eeq

In all cases considered above
\beq
\lim_{N\to\infty}\vstar(r,s) = \frac{\delta_{r,s}}{\pi\left[r^2-\tfrac{1}{4}\right]}\,.
\label{vstarsi11}
\eeq
This says that in the limit $N\to\infty$ each height variable on a given site $(i,j)$  interacts 
only with the height variables on the diagonal $(i+r,j+r)$ passing through
that site
in the direction $(1,1)$, and we recover the large distance behavior of the
potential (\ref{VstarR}) on the
one-dimensional chain of length $MN$ in the limit $MN\to\infty$.

\subsection{Self-dual height-height correlation at $\betastar$}

Let $h_{\veci+\vecr}-h_{\veci}$  be the difference between two 
height variables at sites  $\veci$ and $\veci+\vecr$ in either  dimension $1$ or $2$.
By symmetry we have that
$\langle h_{\veci+\vecr}-h_{\veci}\rangle_\beta=0$. However  the correlation
\beq
\corr(\vecr;\beta)\equiv\langle(h_{\veci+\vecr}-h_{\veci})^2 \rangle_\beta 
\eeq
is a nonvanishing and interesting function of $\vecr$.

 For the DG model on a ring it was shown in reference \cite{KjaerHilhorst82} that, although the correlation $\Corr(R;\beta)\equiv \langle (h_{i+R}-h_i)^2 \rangle_\beta$ is not known  for a generic potential $V(R)$ at any inverse temperature $\beta$, the duality relation (\ref{DualityZring})  for the partition functions implies that  this correlation can be explicitly determined in the case of the self-dual potential  $\Vstar(R)$ at the dual temperature $\betastar=\pi$ defined after (\ref{SelfDualityZtorus}). It reads
\beq
\Corr^\star(R;\betastar)=-\frac{1}{2\pi} \Ustar(R),
\label{corrstarustarRelbis}
\eeq
where the superscript $\star$ of the correlation $G$ signals a  statistical
average with the potential $\Vstar(R)$ and where $\Ustar(R)$ is the periodic potential
associated with $\Vstar(R)$ by (\ref{tVUquad})
and which vanishes at $R=0$.
Relation  (\ref{tVUquad}) can be  seen as a finite difference equation to be solved  for $R$ in the set $\{0,1,\ldots,MN\}$ with the boundary conditions $\Ustar(0)=\Ustar(NM)=0$. 
By rewriting expression (\ref{VstarR}) for $\Vstar(R)$ as a difference of
cotangents with arguments proportional to $R+1$ and $R$ we find that for 
$R=0,1,\ldots,NM$ 
\beq
 \Ustar(R)= -\frac{1}{NM} \sum_{R'=1}^{R} \cot\frac{\pi}{NM}\left(R'-\frac{1}{2}\right),
 \label{UstarRExpressionRPositive}
\eeq
with the understanding that for $R=0$ the sum is empty.
The expression for $\Ustar(R)$ when $R=-NM,-NM+1,\ldots,-1,0$ is obtained by
using the periodicity property $\Ustar(-\vert R \vert)=\Ustar(NM-\vert R
\vert)$ derived from (\ref{UDef}) and rewriting the sum for $\Ustar(NM-\vert R
\vert)$ by taking into account the value $\Ustar(NM)=0$. The result is that
for $R=-NM,-NM+1,\ldots, NM$ we have
\beq
\Ustar(R)= - \frac{1}{NM} \sum_{R'=1}^{\vert R \vert} \cot\frac{\pi}{NM}\left(R'-\frac{1}{2}\right).
\label{UstarRExpression}
\eeq

For the DG model on a  torus an argument  similar  to that presented in
reference \cite{KjaerHilhorst82} shows that, for the potential
$\vstar(r,s)=\Vstar(R)$ at the inverse dual temperature $\betastar$, the correlation  $\corr^\star(r,s;\betastar)= \langle (h_{i+r,j+s}-h_{i,j})^2 \rangle^\star_{\betastar}$ 
can be determined as
\beq
 \corr^\star(r,s;\betastar)=-\frac{1}{2\pi} \ustar(r,s)
 \label{corrstarustarRel}
\eeq 
in which
\beq
\ustar(r,s)=\Ustar(R).
\label{ReluU}
\eeq
Since for the model on the ring $\Ustar(R)$ is known, 
equations (\ref{corrstarustarRel}) and (\ref{ReluU})
allow us to determine
the explicit expression for $\corr^\star(r,s;\betastar)$  on the torus.
This will be the subject of the next subsection.

\subsection{Height-height correlation in the thermodynamic limit for  $N=M+1$}
\label{sec:inflattice}

In the present section we consider the \textit{thermodynamic limit}  where $N=M+1$ and $N$ goes to infinity. Then $M$ and $N$ are coprime, $a=1$, and \break
$R=r+N(s-r)$. Before taking the limit we consider the variables $r$ and $s$ in intervals centered at $(0,0)$. If, for instance, $M$ is even, the intervals read
\beq
- \frac{M}{2} \leq r \leq \frac{M}{2}
\qquad\textrm{and}\qquad - \frac{M}{2} < s \leq \frac{M}{2}.
\label{CenteredIntervals}
\eeq

\subsubsection{Fixed coordinate differences}

In the case of $r=s$ fixed we have that $\ustar(r,r)=\Ustar(r)$ where $\Ustar(r)$ is the sum up to $\vert r\vert$  given in 
 (\ref{UstarRExpression}). In  the thermodynamic  limit  the argument of every cotangent in this sum  is at least of order ${\cal O}(1/N^2)$ so that we can replace $\cot x$ by $1/x$ and $\ustar(r,r)$ becomes
\beq
\ustar(r,r)=-\frac{1}{\pi}\sum_{r'=1}^{\vert r \vert}\frac{1}{r'-\frac{1}{2}} +{\cal O}\left(\frac{1}{N^4}\right).
\eeq
Therefore when $r=s$ the correlation given by (\ref{corrstarustarRel})
 is a nonvanishing function in the thermodynamic limit. It is denoted as
 $\corrstarTh(r,r;\betastar)$ and reads
\beq
\corrstarTh(r,r;\betastar)=\frac{1}{2\pi^2} \sum_{r'=1}^{\vert r\vert }\frac{1}{r'-\frac{1}{2}}.
\label{vstarTh}
\eeq
For large $r$ it behaves as
\beq
\corrstarTh(r,r;\betastar) =\frac{1}{2\pi^2} \left[ \ln \vert r\vert + A_0+{\cal O}\left(\frac{1}{r^2} \right)\right]
\label{vstarThLargeR}
\eeq
with $A_0= C+2\ln2$ where $C$ denotes Euler's constant.

In the case $r\neq s$  it is more convenient to make the  change of  variables $(r,s)\mapsto (r,t)$ with $s=r+t$ and to consider 
\beq
\corrstar(r,r+t;\betastar)=-\frac{1}{2\pi}\Ustar(Nt+r).
\eeq
The expression for $\Ustar(Nt+r)$  is the sum given in (\ref{UstarRExpression}) with $M=N-1$  and $\vert t\vert < N$ according to (\ref{CenteredIntervals}).
When $r$ and $s$ are kept fixed while $N$ and $M=N-1$ become very large, $t$ is fixed and $\vert r+Nt\vert \simeq N\vert t \vert$ with $N \leq  N\vert t \vert \ll N^2$.
Therefore  the argument of every cotangent in the sum is at least of order ${\cal O}(1/N)$  and one can  again replace $\cot x$ by $1/x$, while the upper bound of the sum is of order $N$.
As a result in the thermodynamic limit the leading contribution in the correlation $\corrstar(r,r+t;\betastar)$ is the large distance behavior (\ref{vstarThLargeR}) of expression (\ref{vstarTh}), where the argument $r$ is to be replaced by $Nt=N(s-r)$, 
\beq
\corrstar(r,s;\betastar) \underset{r\neq s}{=} \frac{1}{2\pi^2} \left[ \ln (N \vert s-r \vert )+ A_0+o(1)\right],
\label{vstartFixed}
\eeq
in which $o(1)$ denotes a contribution that vanishes in the limit $N\to\infty$. Equation (\ref{vstartFixed}) expresses that 
when $N=M+1$, according to (\ref{vstarAsOffDiagonal}), two height variables  on parallel diagonals have an interaction whose coupling constant decreases with $N$ so that the variance of their difference increases with $N$.

\subsubsection{Coordinate differences scaled with the lattice size}

Whereas in the preceding subsection we investigated the height-height
correlation $\corrstar(r,s;\betastar)=-(1/2\pi) \Ustar(R)$ 
in the regime of fixed $r$ and $s$ with $N=M+1$ and $N\to\infty$,
it is also interesting to study the nature of this correlation at the scale of
the system, that is, for fixed values of
\beq
\xi = \frac{r}{N},  \qquad \eta = \frac{s}{N-1}
\label{gxieta}
\eeq
where, according to (\ref{CenteredIntervals}), $-\frac{1}{2}<  \xi < \frac{1}{2}$, $-\frac{1}{2}< \eta \leq \frac{1}{2}$  and $N\to\infty$. 
Then
\beq
R=(\eta-\xi)N(N-1)\pmod{N(N-1)}
\eeq
with $-1<\eta -\xi <1$.
Equation (\ref{UstarRExpression}) now leads to
\bea
\Ustar\left((\eta-\xi)N(N-1)\right)
 &=& -\frac{1}{N(N-1)} 
\sum_{R^\prime=1}^{\vert \eta-\xi\vert N(N-1)}
\cot \frac{\pi (R^\prime - \thalf)}{N(N-1)}, \nonumber\\[2mm]
&\equiv&\overline{\Ustar}(\eta-\xi).
\label{Gxieta}
\eea 
with $\overline{\Ustar}(0)=0$ according to (\ref{UstarRExpressionRPositive}).
Since only the absolute value $\vert \eta-\xi  \vert$ appears in the upper
limit of the sum in (\ref{Gxieta}),  it suffices to calculate 
 $\overline{\Ustar}(\eta-\xi)$ with  $0<\eta-\xi<1$. Moreover, according to  expression (\ref{UDef}) for $\Ustar(R)$ as an inverse Fourier transform, and as can be checked on its explicit $R$-dependence given in (\ref{UstarRExpression}), $\Ustar(R)$ has the symmetry  $\Ustar(N(N-1)-R)=\Ustar(R)$. Therefore 
$\overline{\Ustar}(\eta-\xi)$ takes the same value for $\eta-\xi$ and $1-(\eta-\xi)$ and we may further restrict
ourselves to $0<\eta-\xi<1/2$, which we shall do now.

With the present scaling, when $0<\eta-\xi<1/2$,
the argument \break
$\pi(R^\prime-\thalf)/(N(N-1))$ of the cotangent increment in the sum  runs up to values of order $\pi/2$ and for every $R'$
all terms in the large-$N(N-1)$ expansion of the cotangent contribute.  Therefore we shall write $\Ustar= I_0 + I_1$, where $I_0$ and $I_1$ are the sums of the 
contributions of the first term and of all remaining terms, respectively,
in the full expansion. This gives
\bea
I_0 &=&- \frac{1}{\pi} \sum_{R^\prime=1}^{(\eta-\xi) N(N-1)}\frac{1}{R^\prime-\thalf}
\nonumber\\[2mm]
&=& -\frac{1}{\pi}\left[\ln\Big ((\eta-\xi) N(N-1) \Big) +A_0\right]
+ {\cal O}\left( \frac{1}{N^4}\right),
\label{I0}
\eea
where we have used (\ref{vstarThLargeR}),
and
\bea
I_1 &=& -\frac{1}{N(N-1)} \sum_{R^\prime=1}^{(\eta-\xi)N(N-1)}
\Big[ \cot \frac{\pi (R^\prime - \thalf)}{N(N-1)} 
      -\frac{N(N-1)}{\pi (R^\prime - \thalf)}  \Big]
\nonumber\\[2mm]
&=&- \frac{1}{\pi}\int_0^{(\eta-\xi) \pi} \dd u\,\Big[ \cot u - \frac{1}{u}  \Big]
+ o(1)
\nonumber\\[2mm]
&=& -\frac{1}{\pi} \ln\frac{\sin\big((\eta-\xi)\pi \big)}{(\eta-\xi)\pi}
+ o(1).
\label{I1}
\eea 
We obtain $\Ustar$ by adding (\ref{I0}) to (\ref{I1}). When doing so,
a factor $ \eta-\xi $ in the argument of the logarithm cancels against its
inverse, so that the only dependence on $\eta-\xi$ occurs through
$\sin\big( (\eta-\xi) \pi \big)$. We have assumed \break $0< \eta-\xi  <\thalf$, but
as already noticed
$\overline{\Ustar}(\eta-\xi)=\overline{\Ustar}(1-(\eta-\xi))=\overline{\Ustar}(\vert\eta-\xi\vert)$. By
using $\sin (1-\alpha)\pi =\sin \alpha \pi$ we arrive at the result 
\beq
\corrstar(\xi N,\eta (N-1);\betastar) 
\underset{\xi\neq\eta}{=} \frac{1}{2\pi^2}\Big[
\ln \left( N(N-1)\frac{ \sin\big(\vert\eta-\xi \vert\pi\big)}{\pi}\right) 
+  A_0 + o(1) \Big],
\label{Gxietaresult1}
\eeq
valid for all  $-1<\eta-\xi<1$ except $\eta-\xi =0$, that is, for all $(\xi,\eta) \in ]-\frac{1}{2},\frac{1}{2}[ \times ]-\frac{1}{2},\frac{1}{2}]$ except the values $\eta=\xi$.
It so happens that if in (\ref{Gxietaresult1}) we
put again $\xi=r/N$ and $\eta=s/(N-1)$, and expand the resulting expression in
powers of $N$, now at $r$ and $s$ fixed,
we obtain equation (\ref{vstartFixed}).

\section{{Two-dimensional DG model with nearest-neighbor interaction }}
\label{sec:nnbpotential}

In this section we consider the standard DG model with homogeneous isotropic 
nearest-neighbor interaction $\vnnb(r,s)$ on the $N\times M$ torus,
that is,
\beq
\vnnb(r,s) =   J \left[ \big(\delta_{r,-1} + \delta_{r,1}\big) \, \delta_{s,0}
         + \delta_{r,0}  \, \big(\delta_{s,-1} + \delta_{s,1}\big)\right].
\label{xvnn}
\eeq
In this case we do not have a simple formula for the height-height correlation even at a specific temperature
and we shall therefore limit ourselves to studying the dual potential.

\subsection{Dual potential on the $N\times M$ torus}

The Fourier transform of the nearest-neighbor interaction (\ref{xvnn}) reads
\beq
\hvnnb(\kun,\kdeux)=2 J\left[ \cos \kun+ \cos \kdeux \right].
\eeq
The Fourier transform 
of the corresponding dual potential is readily found by means of the general
relation (\ref{tvFT}),
\beq
\htvnnb(\kun,\kdeux)= -\frac{1}{J} \times
\frac{1-\cos(\kun+\kdeux) }{2-\cos\kun-\cos\kdeux}.
\label{FTtvnnb}
\eeq
where we have set $\htvnnb(0,0)=0$.

The two-dimensional lattice Laplacian of a function $f(r,s)$ is defined as
\beq
\Delta_2 f(r,s)= f(r+1,s)+f(r-1,s)+f(r,s+1)+f(r,s-1)-4f(r),
\eeq
and its Fourier transform reads
\beq
\widehat{ \Delta_2 f}(\kun,\kdeux)=-2\left[ 2-\cos\kun-\cos \kdeux\right] \widehat{f}(\kun,\kdeux).
\eeq
Let us now consider the  2D lattice  Coulomb potential with toroidal
periodicity created by a neutral charge distribution $\rho(r,s)$, 
that is, the solution $\Uc_{[\rho]}(r,s)$ of the Poisson equation
\beq
\Delta_2 \Uc_{[\rho]}(r,s) =- \rho(r,s).
\label{PoissonEquation}
\eeq
It has the Fourier transform
\beq
\widehat{\Uc_{[\rho]}}(\kun,\kdeux)=\frac{\widehat{\rho}(\kun,\kdeux)}
{2[2-\cos\kun-\cos\kdeux]}.
\eeq
By comparing this expression with (\ref{FTtvnnb})  and by identifying \break
$2[\cos(\kun+\kdeux)-1]$ as the Fourier transform  of 
\beq
\rho_\textrm{quad}(r,s)= \delta_{r,1}\delta_{s,1}+ \delta_{r,-1}\delta_{s,-1} -2\, \delta_{r,0}\delta_{s,0}\, ,
\eeq
we interpret   $J \times \tvnnb(r,s)$ as the two-dimensional  lattice Coulomb potential created by the quadrupolar charge distribution $(1,-2,1)$ located at sites $(-1,-1)$, $(0,0)$, and $(1,1)$,
respectively.  In other words
\beq
\tvnnb(r,s)=\frac{1}{J}\left[\uc(r+1,s+1)+ \uc(r-1,s-1)-2\uc(r,s)\right],
\label{tvquadripolaire}
\eeq
 where
$\uc(r,s)$ denotes the periodic 2D Coulomb potential created by the neutral
distribution of a single unit charge at the origin and a negative uniform
background with charge $-1/(MN)$ at each site. A priori the solution of the lattice Poisson equation
(\ref{PoissonEquation}) is defined up to an additive constant. The potential
$\uc(r,s)$  is chosen to vanish at the origin and reads
\beq
\uc(r,s)=-\frac{1}{NM}\sum_{(\kun,\kdeux)\neq(0,0)}
\frac{1-\cos(\kun r) \cos(\kdeux s)}{2[2-\cos\kun-\cos\kdeux]}.
\label{uCrsbis}
\eeq
When substituted in (\ref{tvquadripolaire}) this expression yields the
interaction $\tvnnb(r,s)$ dual to the nearest neighbor interaction 
(\ref{xvnn}).

\subsection{Dual potential in the thermodynamic limit for $N=M+1$}

We are now interested in the large-distance behavior of the quadrupolar
potential (\ref{tvquadripolaire}).
In the thermodynamic limit, where $N=M+1$ and $N$ goes to infinity with  $r$ and $s$ fixed, 
the Coulomb potential $\uc(r,s)$ of equation (\ref{uCrsbis}) 
tends to a function $\uc_{\infty}(r,s)$ still given by the same expression
(\ref{uCrsbis}) but with the sums replaced with the appropriate integrals.
Next, we expand $\uc_{\infty}(r,s)$ for large $r$ and $s$ and
obtain \cite{Spitzer64,Uchiyama98}
\beq
\uc_{\infty}(r,s) = -\frac{1}{2\pi}\ln\sqrt{r^2 +s^2} + \mbox{cst} 
+ {\cal{O}}\left( \frac{1}{r^2+s^2} \right).
\label{uClarger}
\eeq
When (\ref{uClarger}) is substituted in (\ref{tvquadripolaire}),
the constant cancels out on the RHS and the result is 
\beq
\tvnnbTh(r,s)\underset{\sqrt{r^2+s^2}\gg1}{\simeq}  \frac{2}{\pi J} \frac{rs}{(r^2+s^2)^2}\,.
\label{tvnbbinftyDecay}
\eeq
We may still set $r=X\cos\phi$ and $s=X\sin\phi$, after which expression (\ref{tvnbbinftyDecay}) becomes
\beq
\tvnnbTh(X \cos \phi, X \sin \phi) \simeq
\frac{\sin 2\phi}{\pi J X^2}\,, \qquad X\to\infty,
\label{tvquadinftySquare}
\eeq
where the factor $\sin 2\phi$ in the numerator brings out the quadrupolar
character of the interaction.

\section{Conclusion}
\label{sec:Conclusion}

We have constructed a new duality for the Discrete Gaussian model on a torus
with arbitrary translation-invariant interactions. 
The duality inverts the temperature and
the interactions are in general anisotropic.
There is a self-dual interaction potential which
we have studied in particular at its self-dual temperature.
We have also considered
the well-known DG model with isotropic nearest-neighbor interactions.
Our work is exact for an $N\times M$ torus with 
finite $N$ and $M$ which should be coprime.
This condition has its origin in the Chinese Remainder Theorem,
which we invoke to transpose known one-dimensional results to the
two-dimensional torus.
The mapping avoids the appearance of any kind of seam on the torus.
One simple way to satisfy the coprime condition is to set $M=N-1$, where
$N$ is an arbitrary integer. At several points in our discussion we have 
taken the thermodynamic limit $N\to\infty$.
Another similar duality can be derived for a neutral charge system
corresponding to the Discrete Gaussian model and will be discussed elsewhere.

We have not in this paper attempted to be fully general.
Indeed the same method may be used to construct dualities in arbitrary dimension $d$
on a hypertorus of $N_1\times N_2 \times \ldots\times N_d$ sites,
provided the $N_j$ are all mutually coprime.
Moreover this work relates partition functions, hence free energies,
as well as correlation functions, in dual pairs of models. 
In the case of the self-dual potential $\vstar(r,s)$ and at the self-dual inverse temperature $\betastar$ 
the relation allows us to determine the spatial  correlation as discussed in  section \ref{sec:Selfduality}.
The  study of the possible critical regimes  requires  further investigations.
However the present paper contributes to the large body of exact results,
in particular for duality relations, in lattice models.

\end{document}